\documentclass[conference]{IEEEtran}
\IEEEoverridecommandlockouts
\usepackage{amsmath,amssymb,amsfonts}
\usepackage{graphicx}
\usepackage{textcomp}
\usepackage{xcolor}
\usepackage{comment}
\usepackage{algorithm}
\usepackage[noend]{algpseudocode}
\usepackage{xcolor}
\usepackage{fontawesome}
\usepackage{tikz}
\usepackage{comment}
\usepackage{centernot}

\usepackage{subfig,hyperref}
\def\BibTeX{{\rm B\kern-.05em{\sc i\kern-.025em b}\kern-.08em
    T\kern-.1667em\lower.7ex\hbox{E}\kern-.125emX}}
\begin{document}

\title{How Good LLM-Generated Password Policies Are?\thanks{*These authors contributed equally to this work.}\\
}

\author{\IEEEauthorblockN{Vivek Vaidya*}
\IEEEauthorblockA{\textit{Department of Computer Science} \\
\textit{Rutgers University}\\
New Brunswick, USA \\
vivek.vaidya@rutgers.edu}
\and
\IEEEauthorblockN{Aditya Patwardhan*}
\IEEEauthorblockA{\textit{Department of Computer Science} \\
\textit{Stony Brook University}\\
Stony Brook, USA \\
aapatwardhan@cs.stonybrook.edu}
\and
\IEEEauthorblockN{Ashish Kundu}
\IEEEauthorblockA{\textit{Cisco Research} \\
San Jose, USA \\
ashkundu@cisco.com}
}


\maketitle

\begin{abstract}
Generative AI technologies, particularly Large Language Models (LLMs), are rapidly being adopted across industry, academia, and government sectors, owing to their remarkable capabilities in natural language processing. However, despite their strengths, the inconsistency and unpredictability of LLM outputs present substantial challenges, especially in security-critical domains such as access control. One critical issue that emerges prominently is the consistency of LLM-generated responses, which is paramount for ensuring secure and reliable operations.

In this paper, we study the application of LLMs within the context of Cybersecurity Access Control Systems. Specifically, we investigate the consistency and accuracy of LLM-generated password policies, translating natural language prompts into executable \texttt{pwquality.conf} configuration files. Our experimental methodology adopts two distinct approaches: firstly, we utilize pre-trained LLMs to generate configuration files purely from natural language prompts without additional guidance. Secondly, we provide these models with official \texttt{pwquality.conf} documentation to serve as an informative baseline. We systematically assesses the soundness, accuracy, and consistency of these AI-generated configurations. Our findings underscore significant challenges in the current generation of LLMs and contribute valuable insights into refining the deployment of LLMs in Access Control Systems.
\end{abstract}


\section{Introduction}\label{sec:intro}
Access control systems—including robust password policy enforcement—are fundamental to cybersecurity, ensuring that sensitive resources remain accessible only to authorized users. Traditionally, Linux systems have relied on password authentication modules (PAM) and associated files such as pwquality.conf to enforce desired password policy for the quality standards. Large language models (LLMs) such as ChatGPT~\cite{openai2022chatgpt}, Gemini~\cite{geminiteam2024geminifamilyhighlycapable}, have been studied in the context of automation of cybersecurity tasks and operations. In this paper, we are studying the problem of how good the LLM-generated password policies for Linux are especially for Linux PAM?

Recent advances in Large Language Models (LLMs) and AI agents offer promising opportunities to automate the generation of access control policies. In particular, using LLMs to translate text-based password policies into usable \emph{pwquality.conf} files that can directly slot into Linux systems to enforce standards could revolutionize how organizations manage and update their security protocols. However, integrating AI-driven systems into such critical roles demands an incredibly high standard of precision, consistency, and reliability. Even minor inconsistencies in the generated policies could lead to policy engine conflicts within a network due to differences on local machines. 

While prior research has evaluated LLM output consistency in broader cybersecurity contexts using frameworks such as BECEL and TruthEval~\cite{jang-etal-2022-becel, khatun2024truthevaldatasetevaluatellm}, these studies have largely overlooked the specific challenges inherent in password policy control. Password policies, by their nature, require consistency; even subtle deviations in rule enforcement can compromise the security of an entire system. A more recent study ~\cite{patwardhan2024automated} has also concluded that while LLMs have made significant strides in recent times, they are still inconsistent and tend to hallucinate in their more abstract plain text responses. 

In this paper, we studied the problem of how good the LLM-generated password policies for Linux are? We have developed a framework that generates and analyzes these LLM-generated password policies for the following key properties: Consistency, Hallucination,  Correctness, and Incompleteness. 

This paper explores a slightly different side of consistency: that's more focused on directly checking parameter assignments as opposed to judging semantic consistency with natural language responses, to answer the question: how consistent are modern Large Language Models at generating usable password policies based on plain text instructions? 

{\bf Organization of the paper}: Section~\ref{sec:background} and Section~\ref{sec:framework} describe the importance of Access Control Systems, Password Policies, the key definitions for this study, and the motivation behind this study. In Section~\ref{sec:implementation}, we have presented our novel framework for Consistency and Correctness in the context of password configuration files. Section~\ref{sec:results} presents our findings and discusses the impact of these findings on real-world systems. Section~\ref{sec:related} discusses the related work, and Section~\ref{sec:conclusion} concludes the paper with future work.

\section{Background and Problem Motivation} \label{sec:background}
Access Control Systems in Linux establish the overarching guidelines for resource protection by implementing both discretionary and mandatory controls. Traditional Unix file permissions, along with more granular Access Control Lists (ACLs) and mandatory access control systems like SELinux~\cite{love2005selinux} and AppArmor~\cite{zhu2021apparmor}, define how users and processes interact with the system. This general framework not only regulates file and process permissions but also lays the groundwork for authenticating users—a process that directly influences how password management is conducted. 

Password Management is a critical component within this broader access control strategy. It focuses on the secure creation, storage, and periodic updating of user credentials, typically stored in encrypted forms in files like /etc/shadow. Effective password management is crucial because it mitigates risks associated with brute force and guessing attacks, ensuring that only properly authenticated users can gain system access. Thus, robust password management is an indispensable extension of the overall access control system, bridging the gap between user identity and resource access. 

\begin{figure}[tb]
    \centering
    \includegraphics[scale=0.5]{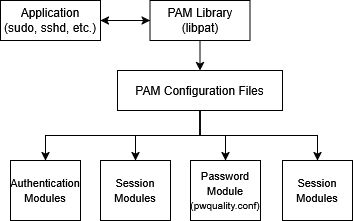} 
    \caption{PAM with configuration files in Linux}
    \label{fig:Original_PAM_config_files}
\end{figure}

To integrate and streamline these authentication processes, Linux employs Pluggable Authentication Modules (PAM). PAM offers a modular framework that decouples authentication mechanisms from application logic, allowing administrators to configure various methods—ranging from traditional passwords to multi-factor authentication—across different services like login, SSH, or sudo. By providing a unified and adaptable approach, PAM not only simplifies authentication but also reinforces the security policies set by the access control and password management systems. Figure \ref{fig:Original_PAM_config_files} describes how the PAM module with its configuration files is stored in Linux.

Supporting all these mechanisms are Password Policies, which set formal guidelines for creating and maintaining strong passwords. These policies—defining criteria such as minimum length, complexity, expiration intervals, and reuse restrictions—are often implemented via PAM modules and system configuration files. Password policies ensure that the password management system operates effectively, directly contributing to the overall robustness of access control. 

In the current landscape of generative AI, large language models (LLMs) are emerging as powerful tools for automatically generating and refining these security policies. By leveraging LLMs, system administrators can efficiently produce up-to-date password policies and other access control configurations that reflect the latest best practices and threat intelligence. This paper examines how LLM-generated policies can be integrated within the Linux File Subsystem. Our study focuses on how these AI-assisted policies can be incorporated into the system by administrators lacking adequate training or experience, which leads to authentication issues or gives rise to vulnerabilities that can then be exploited by malicious users. 

In this paper, we specifically focus on automating the generation of the pwquality.conf file, a configuration-level component that defines the password policy rules enforced by Linux systems via the PAM stack. This file plays a critical role in bridging human-readable policy descriptions with machine-enforced authentication rules. While it does not perform enforcement itself, its parameters are consumed by the pam\_pwquality.so module—a key enforcement mechanism within PAM. If the pwquality.conf file is missing, misconfigured, or contains outdated parameters, the PAM module silently falls back on its internal hardcoded defaults, referred to in this paper as the “failsafe” configuration. This fallback behavior, although intended to preserve system integrity, can unintentionally bypass stricter security policies if the configuration is inconsistent. Therefore, ensuring that LLM-generated pwquality.conf files are correct, complete, and aligned with best practices is vital for maintaining effective access control. Our work serves as a first step in automating configuration generation at this interface, with methods that can be extended to other security-critical files in the Linux ecosystem, such as login.defs or sshd\_config.

\section{Automated Generation of Policies by LLMs} \label{sec:framework}

In this study, we focus on the generation of password configuration files used in conjunction with the Pluggable Authentication Modules (PAM) to configure password quality requirements and enforce password complexity rules. Specifically, we focus on the generation of the pwquality.conf file using Large Language Models. 

\subsection{Key Properties}
Some key properties in the context of the consistent generation of configuration files are as follows:

\subsubsection{Consistency}\label{sec:subsection-Consistency}
Consistency in the context of Large Language Models refers to an LLM's ability to generate similar responses when given the same or semantically identical prompts. Although past studies such as \cite{patwardhan2024automated} have explored abstract forms of consistency, like semantic consistency, in this study, we adopt a more pragmatic definition: an LLM is considered consistent if it generates configuration files that behave exactly the same way. This definition is particularly critical in the realm of password policies, where consistency across a network is essential for uniformly enforcing security standards. Algorithm \ref{alg:comparison} provides a systematic method for assessing consistency by quantifying the similarity between two configuration files based on their parameter assignments. Specifically, the algorithm computes a similarity metric that quantifies the number of parameters in one generated file that match the corresponding parameters in another generated file, offering a quantitative measure of consistency. 

\subsubsection*{Inconsistency}
As opposed to consistency, we define an LLM as an Inconsistent LLM if it generates two configuration files with different parameters for the same set of natural language policies. Such a situation may lead to inconsistencies between systems. If an LLM interprets the same natural language password policy in more than one way, it could create discrepancies on different machines, leading to inconsistent access control, potential security breaches, and authentication errors. Inconsistency is closely intertwined with Hallucinations of parameters in the generated files. When an LLM fabricates or misinterprets parameters, the resulting configuration files may diverge from the expected behaviour, undermining the reliability of Access Control Systems.

\subsubsection{Correctness}\label{sec:subsection-Correctness}
We define correctness using a manually generated gold-standard benchmark. We implement this by defining sets of natural language policies~\cite{purplesec_passwords, lumificyber_passwords, securden_passwords, nist_80063b, cis_password_guide, linux_password_defaults} to serve as prompts to the LLM. For each set of natural language policies, we manually define a pwquality.conf file to serve as a benchmark, i.e., a 'correct' answer. This benchmark file enforces the natural language policy to the extent of what is possible to set in the pwquality.conf file, such as setting a password's minimum length and complexity requirements, while ignoring policies outside the scope of the file, such as requiring rate limitation to be implemented externally. LLMs are then prompted with these natural language policy descriptions, and they generate their own configuration files, intended to implement the specified policies. To measure the accuracy of these generated files, we use our Correctness algorithm (Algorithm \ref{alg:corr}) that operates on a parameter-by-parameter basis. Specifically, the algorithm checks each parameter in the LLM-generated file against the corresponding parameter of the benchmark file. 

\subsubsection{Hallucination}

Hallucinations~\cite{mcdonald2024reducing} refer to instances where an LLM fabricates or misinterprets configuration parameters, resulting in deviations from the documented defaults. We define Hallucination as discrepancies that are created when an LLM produces outputs with varying parameter assignments or unexpected deviations from the values prescribed in the official pwquality.conf documentation. For example, if an LLM introduces a parameter that does not exist in the benchmark or assigns a value that contradicts the documented default (such as assigning an incorrect minimum length), such behavior indicates that the model is generating extraneous or erroneous information. These hallucinated outputs cause the resulting configuration files to be diverted from their expected behavior, thus undermining the reliability and security of access control mechanisms they are meant to enforce. Even if an output is semantically consistent, any deviation from the official parameters results in inconsistent policy enforcement across the network. By evaluating the generated files on a parameter-by-parameter basis, we can quantify the degree of hallucination. This not only measures the consistency of LLM responses but also provides insights into the impact of hallucination on the overall trustworthiness of LLMs. To quantify the extent of hallucinations in the generated configuration files, we employ our Hallucination Analysis Algorithm (Algorithm \ref{alg:comparison}). It identifies and counts parameters that are fabricated. The algorithm then computes the average number of hallucinated parameters per file, providing a dual measure of both consistency and the incidence of hallucination.

\subsubsection{Incompleteness}
The property of incompleteness refers to instances where an LLM-generated configuration file does not fully specify all the parameters required to enforce a given password policy. This can occur when the LLM either omits certain parameters entirely or provides only partial information about them, leading to outputs that may lack comprehensive coverage of the intended security settings. In our framework, we address this issue by interpreting missing parameters as being implicitly assigned their default values as documented in the official pwquality.conf guidelines~\cite{arch_pwquality}. However, incompleteness still represents a significant challenge, as it can indicate that the LLM's understanding of the natural language prompt is partial or that the model is unable to capture all the nuances of the policy.

\subsection{Analysis of Properties}

In our framework, we evaluate LLM-generated configuration files using the four key properties described above: Consistency, Correctness, Hallucination, and Incompleteness. Consistency refers to the LLM’s ability to produce functionally equivalent outputs when presented with the same or semantically identical prompts, ensuring uniform enforcement of access control policies. Correctness is established by comparing these outputs against a manually defined gold-standard benchmark, which guarantees that the intended password policies are accurately implemented. Hallucination captures instances where the LLM fabricates or misinterprets parameters, leading to deviations from documented defaults and potentially introducing security vulnerabilities. Incompleteness arises when the LLM fails to specify all required parameters, leaving gaps that might compromise comprehensive policy enforcement. As Table \ref{tab:impact} shows, collectively, these properties directly influence critical aspects such as security, compliance, usability, and Linux PAM validation. We therefore define an LLM-generated configuration as sound if it meets all four criteria—delivering outputs that are consistent, correct, free of hallucinations, and complete. This final check for soundness serves as an aggregate measure of the overall trustworthiness and reliability of the generated configurations in secure access control systems.

\definecolor{headergray}{gray}{0.9}
\definecolor{rowgray}{gray}{0.95}

\begin{table}[ht]
\small
\setlength{\tabcolsep}{3pt}
\centering
\caption{Impact of LLM-Generated Config Properties}
\label{tab:impact}
\begin{tabular}{|p{1.9cm}|p{1.4cm}|p{1.4cm}|p{1.4cm}|p{1.6cm}|}
\hline
\textbf{Property} & \textbf{Security} & \textbf{Compliance} & \textbf{Usability} & \textbf{PAM Validation} \\
\hline
Consistency    & High: Fewer vulnerabilities & High: Aligns with policies & Moderate: Predictable UX & High: Better PAM acceptance \\
\hline
Correctness    & High: Blocks bad access & High: Meets standards & Moderate: Fewer errors & High: Passes PAM checks \\
\hline
Hallucination  & Low: Fake entries risky & Low: Can break rules & Low: Confuses users & Low: May be rejected \\
\hline
Incompleteness & Low: Gaps reduce safety & Low: May miss reqs & Low: Causes instability & Low: Likely rejected \\
\hline
\end{tabular}
\end{table}

\subsection{Compliance in LLM-Generated Policies}

Leveraging LLM-generated policies introduces additional layers of complexity to the compliance landscape. Since LLMs are trained on diverse datasets, they may inadvertently incorporate outdated or misaligned practices, thereby necessitating meticulous review processes to ensure adherence to established legal, regulatory, and industry standards. Robust validation mechanisms are therefore essential to maintain consistency as regulatory requirements evolve. 

To address these challenges, we propose a definition of soundness for LLM-generated policies. We define an LLM as sound if it generates functionally equivalent files. For example, consider the pwquality.conf file, which automatically applies default settings when certain parameters are omitted, in accordance with the Linux documentation~\cite{arch_pwquality}. In this context, explicitly defining a parameter such as minlen=8 is functionally identical to omitting it altogether. Consequently, if an LLM produces two configuration files—one where minlen=8 is explicitly stated and one where the parameter is omitted—the outputs are deemed sound. 

Nonetheless, despite the implementation of robust validation mechanisms, compliance issues may still arise in real-world scenarios. For instance, if a password policy description fails to specify the default value for the parameter controlling the minimum days after which a password should be changed (e.g., pass\_max\_days), the LLM might generate two different configuration files, one assigning a value of 90 and the other 180. Such discrepancies, stemming from both inconsistency and hallucination, underscore the critical need for a thorough analysis of Generative AI-based applications in Access Control Policies.

\section{LLM-Generated PAM Password Policies and  Soundness Analysis} \label{sec:implementation}
To address the aforementioned issue, we propose two experimental frameworks. In the first, we assess pre-trained LLMs - here, the model generates a pwquality.conf file based solely on its inherent training, and the output is compared against other LLM-generated files and against our benchmark file. In the second framework, we provide the model with the official documentation \ref{subsubsec:doc inclusion} along with a prompt to generate the pwquality.conf file, and similar to the first, we compare the result to other LLM-generated versions and to the benchmark file. This dual-framework approach enables us to investigate the impact of consistency, Hallucination, Correctness, and Incompleteness on the Soundness of the generated configuration files. 

\subsection{System Design}

\begin{figure}[tb]
    \centering
    \includegraphics[width=\linewidth]{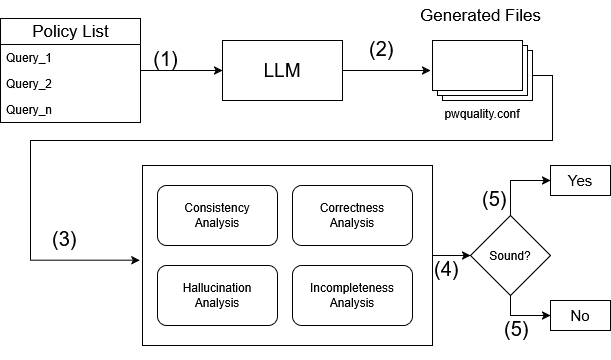}
    \caption{Password Policy Generation and HICC Analysis}
    \label{fig:Policy_File_Generation_and_Usage}
\end{figure}

Figure \ref{fig:Policy_File_Generation_and_Usage} outlines our system, where an LLM generates the pwquality.conf file based on our policies and replaces the Linux pwquality.conf file and passes through the Correctness Algorithm (Algorithm \ref{alg:corr}) and the Consistency and Hallucination Analysis (Algorithm \ref{alg:comparison}). This process ensures that the configuration is both syntactically and semantically sound before being organized into the standard PAM module, integrating automated generation with verification to produce a policy-compliant configuration file. 

We define an LLM-generated configuration as sound if it meets all four criteria: it consistently produces functionally equivalent outputs, accurately reflects the intended password policy (correctness), does not introduce fabricated or erroneous parameters (absence of hallucination), and fully specifies the required configuration parameters (completeness). This final check for soundness serves as an aggregate measure of the trustworthiness of the generated file. 

\subsubsection{Augmentation}\label{subsubsec:doc inclusion}
In our study, we investigate the impact of this approach on both \ref{sec:subsection-Consistency} and \ref{sec:subsection-Correctness}.  It is important to note that our approach does not incorporate Retrieval Augmented Generation (RAG) or In-Context Learning. Our work only leverages the pre-trained natural language generation of LLMs, without additional fine-tuning or task-specific adaptation. We supply the LLM with the link to the official documentation of the pwquality.conf file~\cite{arch_pwquality} that details all relevant keywords and all proper assignment rules for each parameter. By embedding this documentation into our prompt, we enable the LLMs to reference a trusted source and ground their output. This method serves as a form of implicit training: the LLM can leverage the provided context to better interpret natural language descriptions of password policies and translate them into configuration files that align with the documented defaults and recommendations. 

It is important to clarify that our framework does not attempt to modify or interact directly with the internal logic of the PAM modules or their failsafe defaults. Instead, our automation is restricted to the generation of the pwquality.conf file, which is interpreted at runtime by the pam\_pwquality.so enforcement module. If this file is absent, malformed, or uses deprecated parameters, the PAM system will default to internally hardcoded policies to prevent misconfiguration-induced vulnerabilities. As such, our evaluation assumes the presence of a valid pwquality.conf file and aims to measure how well LLMs can produce configurations that remain within the boundaries of what the PAM system recognizes and enforces. This approach ensures practical relevance while avoiding unintended interference with PAM’s failsafe mechanisms.

\subsection{Generation of pwquality.conf}

Algorithm \ref{alg:avg_consis and param gen} is our file generation algorithm, which returns the average scores for a certain prompt for an LLM. For each LLM, it generates a defined number of responses, which we have set to five, and stores them. Then it calls Algorithm \ref{alg:comparison} for each pair of responses and returns the average number of hallucinated responses, consistency including hallucinated parameters, and consistency not including hallucinated parameters.

\begin{algorithm}[tb]
\caption{Parameter Generation}
\label{alg:avg_consis and param gen}
\begin{algorithmic}[1]
\Procedure{Avg\_Consistency}{\textit{LLM, prompt, iterations}}

\State $resps \gets []$

\For{$i \in 0 \dots iterations$}
    \State $resps[i] \gets gen\_resp(LLM, prompt)$
\EndFor
\State $total \gets []$

\For{$i \in 0 \dots iterations$}
    \For{$j \in i \dots iterations$}
        \State $total += Response\_Comparison(resps[i],\allowbreak\ resps[j])$
    \EndFor
\EndFor
\State $sum \gets iterations*((iterations-1)/2)$

\Return $[total[0]/sum, total[1]/sum, total[2]/sum]$
\EndProcedure
\end{algorithmic}
\end{algorithm}

\subsection{Consistency and Hallucination Analysis}

Algorithm \ref{alg:comparison} is our Consistency and Hallucination Analysis algorithm for a single pair of files. It loads the defaults and then reads each response from both responses to fill up two hashmaps, with keys being the parameter, and values being the assignment. The algorithm then compares each one. If the parameter's assignments are the same, num\_same, which keeps track of the number of parameters that are assigned the same, is incremented. If the parameter is a real parameter, then num\_same\_real, which keeps track of the number of real parameters that are assigned the same, is updated. If it isn't num\_hal, which keeps track of the number of hallucinated parameters is updated. Afterwards, it returns an array containing the average number of hallucinated parameters in each file (num\_hal/2), and the average number of parameters assigned to the same value, both including hallucinated parameters and not including hallucinated parameters.

\begin{algorithm}[tb]
\caption{Hallucination and Consistency Analysis}
\label{alg:comparison}
\begin{algorithmic}[1]
\Statex \textbf{Input:} \textit{Response1} - One response from an LLM 
\Statex \textbf{Input:} \textit{Response2} - Another response from an LLM for the same prompt generated at a different time
\Procedure{Response\_Comparison}{\textit{Response1, Response2}}
\State $def\_params \gets load\_defaults()$
\State $resps1 \gets def\_params.copy()$
\State $resps2 \gets def\_params.copy()$

\For{$param, val \in response1$}
    \State $resps1[param] \gets val$
\EndFor
\For{$param, val \in response2$}
    \State $resps2[param] \gets val$
\EndFor
\State $num\_hal \gets 0$
\State $num\_same \gets 0$
\State $num\_same\_real \gets 0$
\State $all\_keys = [resps1.keys() + resps2.keys()]$

\For{$k \in all\_keys$}
    \If{$resps1[k] == resps2[k]$}
        \State $num\_same = num\_same+1$
        \If{$k \in def\_params$}
            \State $num\_same\_real = num\_same\_real+1$
        \Else
            \State $num\_hal = num\_hal+1$
        \EndIf
    \EndIf
\EndFor
\State $lenreal \gets len(def\_params)$
\State $len \gets len(all\_keys)$

\Return $[num\_hal/2,\allowbreak\ num\_same/len,\allowbreak\ num\_same\_real/lenreal]$
\EndProcedure
\end{algorithmic}
\end{algorithm}

\subsection{Correctness and Incompleteness Analysis}

Algorithm \ref{alg:corr} is our Correctness and Incompleteness Analysis algorithm. Similar to our consistency and hallucination algorithm, it prompts an LLM a certain number of times and stores those responses. Instead of comparing these responses with each other, it compares each one with the provided benchmark, which contains the correct assignment for each parameter. It then returns the average of the percentage of correctly assigned parameters. Additionally, the algorithm identifies incompleteness by checking for any missing parameters and interpreting their absence as an implicit assignment to the documented default values. This dual evaluation enables us to quantify both the correctness and the completeness of the generated configuration files, ensuring that all required parameters are accurately and fully specified.

\begin{algorithm}[tb]
\caption{Correctness and Incompleteness Analysis}
\label{alg:corr}
\begin{algorithmic}[1]
\Procedure{Avg\_Correctness}{\textit{LLM, prompt, iterations, benchmark}}

\State $resps \gets []$

\For{$i \in 0 \dots iterations$}
    \State $resps[i] \gets gen\_resp(LLM, prompt)$
\EndFor
\State $total \gets []$

\For{$i \in 0 \dots iterations$}
    \State $total+=Response\_Comparison(resps[i], benchmark)$
\EndFor
\Return $total[2]/iterations$
\EndProcedure
\end{algorithmic}
\end{algorithm}

\section{Experimental Discussions}\label{sec:results}
\subsection{Outlier LLM Behavior}
There are some things the raw data does not take into account. The algorithm goes parameter by parameter and checks the assignment of each one, but doesn't check the overall state of the file. A few LLMs exhibit outlier behavior that is not measured by the data. 

Bloom~\cite{workshop2023bloom176bparameteropenaccessmultilingual} refuses to generate anything at all, returning only blank files. If this file were plugged into pwquality.conf, it would default to all the defaults for all parameters, hence why it still doesn't score badly on accuracy. Since it always spits out a blank file, it also has a perfect 100\% in consistency. 

Cohere~\cite{cohere2023} occasionally forgets the '\texttt{=}' sign in it's assignments. For example, it will sometimes spit out minlen 8 instead of minlen=8. While this may be the correct assignment, the file doesn't work if plugged in directly. 

Llama3~\cite{touvron2023llamaopenefficientfoundation} includes headers inside brackets in the files it generates, for example, it'll put the cracklib parameter under a [dictionary] header and everything else under a [general] header.

\subsection{Consistency}

\begin{figure}[tb]
    \centering
    \includegraphics[width=\linewidth]{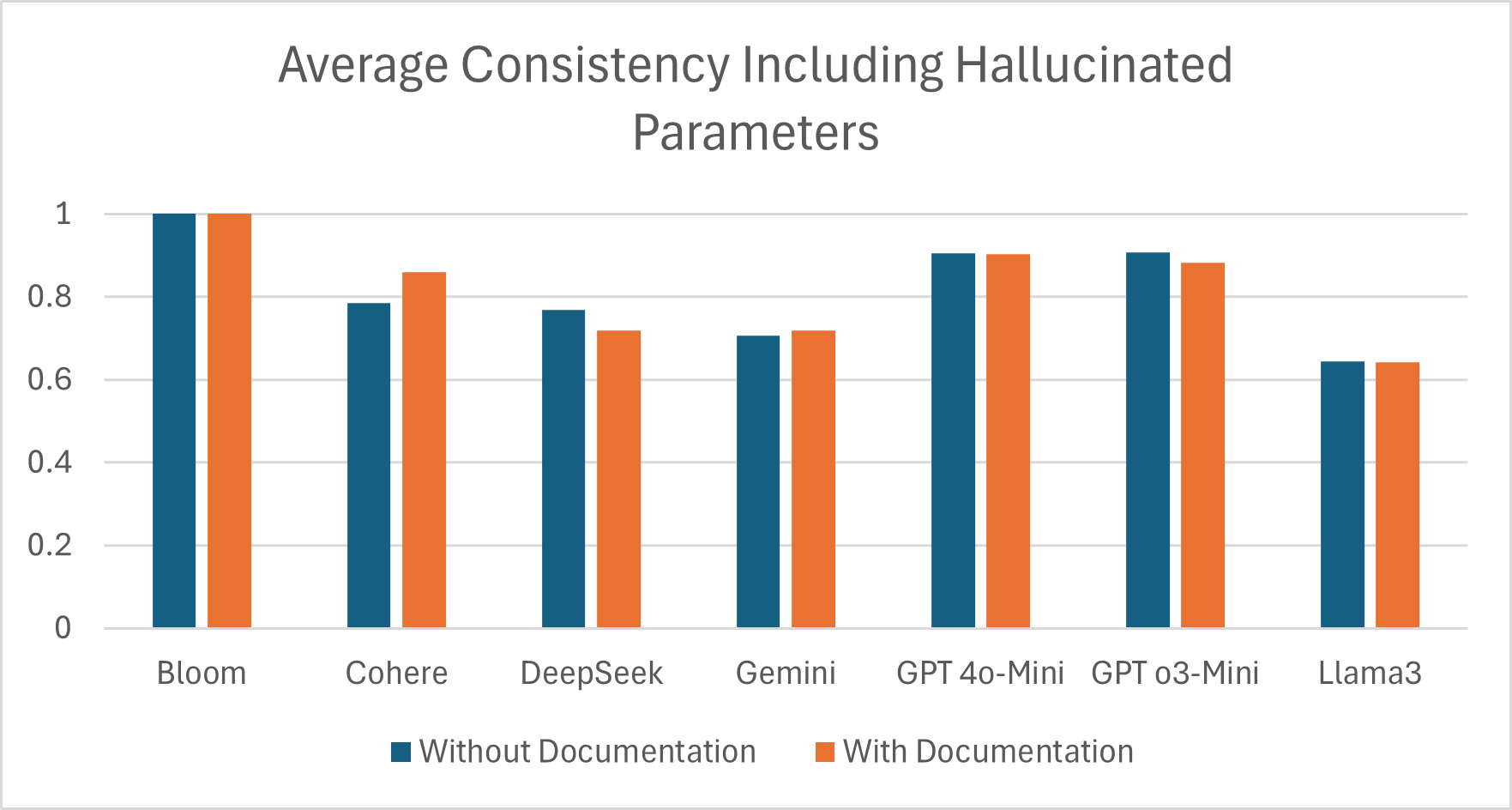}
    \caption{Average LLM Consistency (Including Hallucinated Parameters)}
    \label{fig:avg_cons_hal}
\end{figure}
\begin{figure}[tb]
    \centering
    \includegraphics[width=\linewidth]{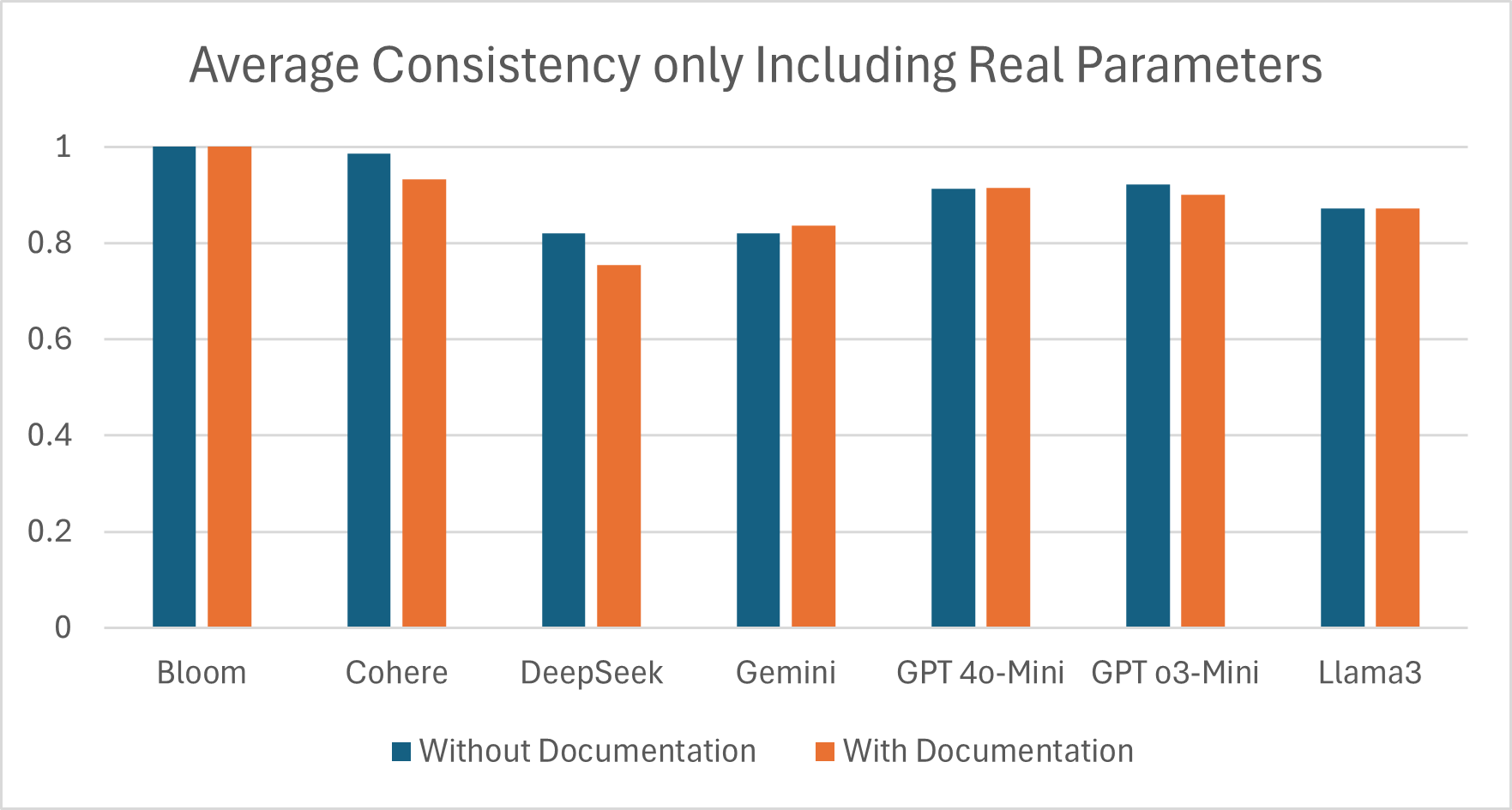}
    \caption{Average LLM Consistency (Only Including Real Parameters)}
    \label{fig:avg_cons_real}
\end{figure}

Figures \ref{fig:avg_cons_hal} and \ref{fig:avg_cons_real} refer to LLM consistency in response generation, including the hallucinated parameters and only including the real parameters, respectively. 

\subsubsection{Consistency Including Hallucinated Parameters}
When including hallucinated parameters as seen in Figure \ref{fig:avg_cons_hal}, the most consistent models excluding Bloom are the two OpenAI models, with GPT 4o-mini~\cite{openai2024gpt4omini} doing slightly better than GPT o3-mini~\cite{openai_o3mini}. Throughout every LLM, including documentation, has close to negligible effects on consistency. o3-mini and DeepSeek~\cite{deepseekai2025deepseekv3technicalreport} display a slight reduction in consistency when documentation is provided, while Cohere and Gemini~\cite{geminiteam2024geminifamilyhighlycapable} display a slight increase. The two models that experience an increase, on average, hallucinate more parameters than the ones that experience a reduction, and both tend to hallucinate more when given documentation. They likely experience this increase because there are more false parameters generated and they are fairly consistent in their hallucinations. 

\subsubsection{Consistency Only Including Real Parameters}
Removing hallucinated parameters is more useful for practically measuring consistency in this study because it more directly measures differences in how the password policies will function in practice. Hallucinated parameters being different won't affect the policy engine at all, because it won't even check them. Real parameters being incorrectly assigned will functionally change the policy. When hallucinated parameters are not included, as seen in Figure \ref{fig:avg_cons_real}, the average consistency for most LLMs seems to increase. Interestingly, trends for the effects of including documentation on consistency remain the same for every LLM except Cohere. Cohere does considerably better when not given the documentation, as opposed to doing better with the documentation when including hallucinated parameters. This is likely because including documentation for whatever reason makes Cohere much more likely to hallucinate, increasing the number of parameters that can be correctly assigned when including them. Interestingly, Cohere is actually the stronger performer here even when including documentation, even outpacing the newer and more refined OpenAI models. Despite being the model that hallucinates the most, Cohere is also the most functionally consistent out of the measured models. 

\subsection{Correctness}

\begin{figure}[tb]
    \centering
    \includegraphics[width=\linewidth]{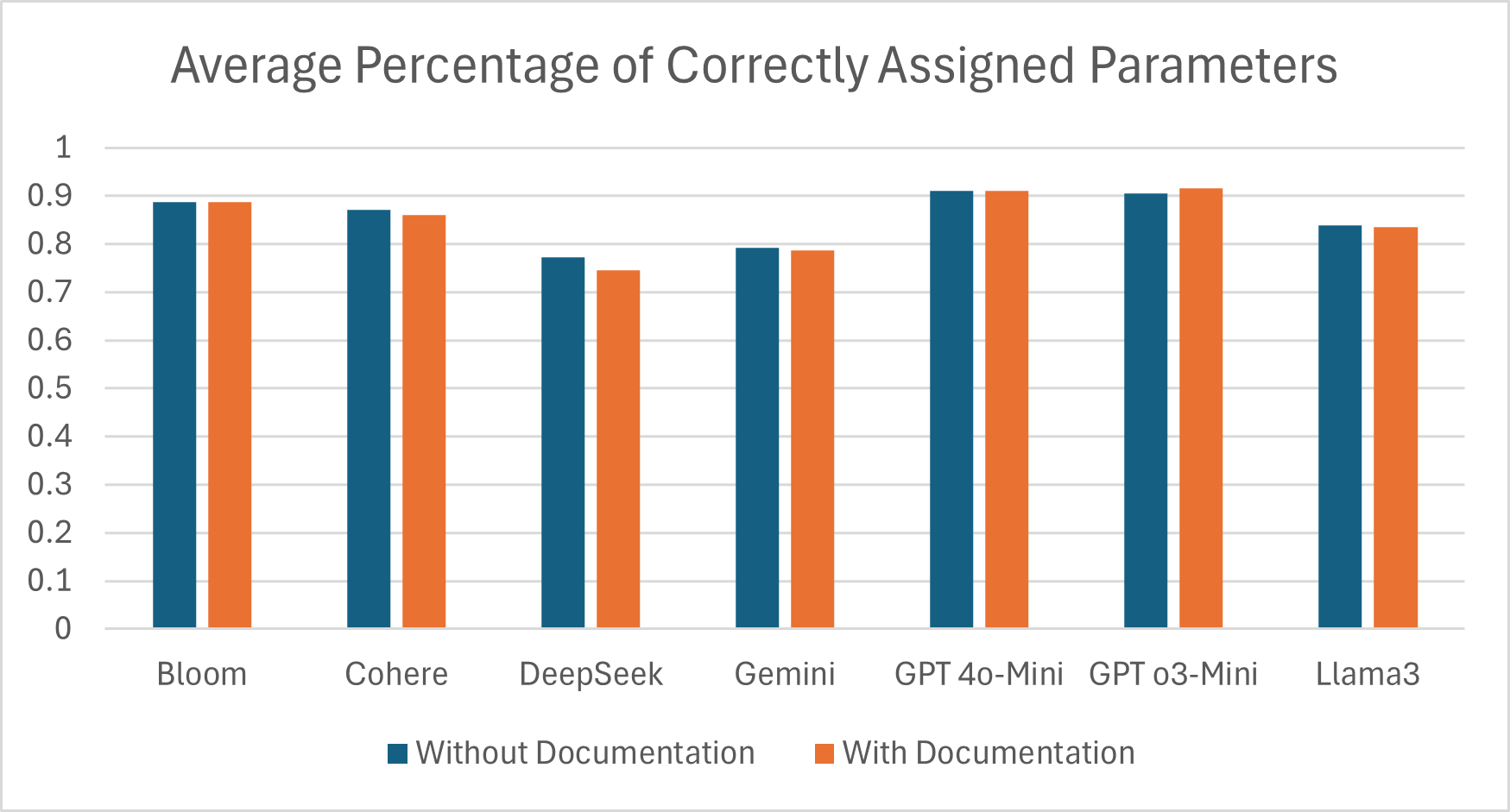}
    \caption{Average LLM Accuracy}
    \label{fig:avg_acc}
\end{figure}

Figure \ref{fig:avg_acc} displays the average number of correctly assigned parameters per response by LLM when compared to our manually created benchmark files for each prompt. Bloom does quite well here with its blank files since the default and a few of our red herring prompts inflate its score. Not including Bloom, the OpenAI models once again perform the best, with o3-mini outscoring 4o-mini both with and without documentation, and increasing when provided with documentation. o3-mini is interestingly the outlier in that quality. Giving documentation to every other LLM either has no effect or is actively detrimental to their accuracy. 

\subsubsection{Effects of Incorrect Parameter Assignments}
If an LLM sets an existing parameter to an invalid value, this will cause an issue, unlike with hallucinated parameters. For example, if you set minlen to a string, when trying to set a new password, the module will return "Authentication token manipulation error" and leave the password unchanged. Next time the command is run, the PAM module will default to the default rules while ignoring anything in the file. For example if I set minlen to a string, and the difok to 2, saying that at there must be at least two characters in the new password that aren't in the old one, this will be ignored in favor of the default, setting difok to 1, just because minlen was set to an invalid value. LLMs have made similar mistakes before; for example, Cohere forgot the equals sign on a few of its assignments on 3/5 of the files generated for a prompt, meaning that 3/5 times the policy wouldn't be enforced at all.

\subsection{Hallucination}
\begin{figure}[tb]
    \centering
    \includegraphics[width=\linewidth]{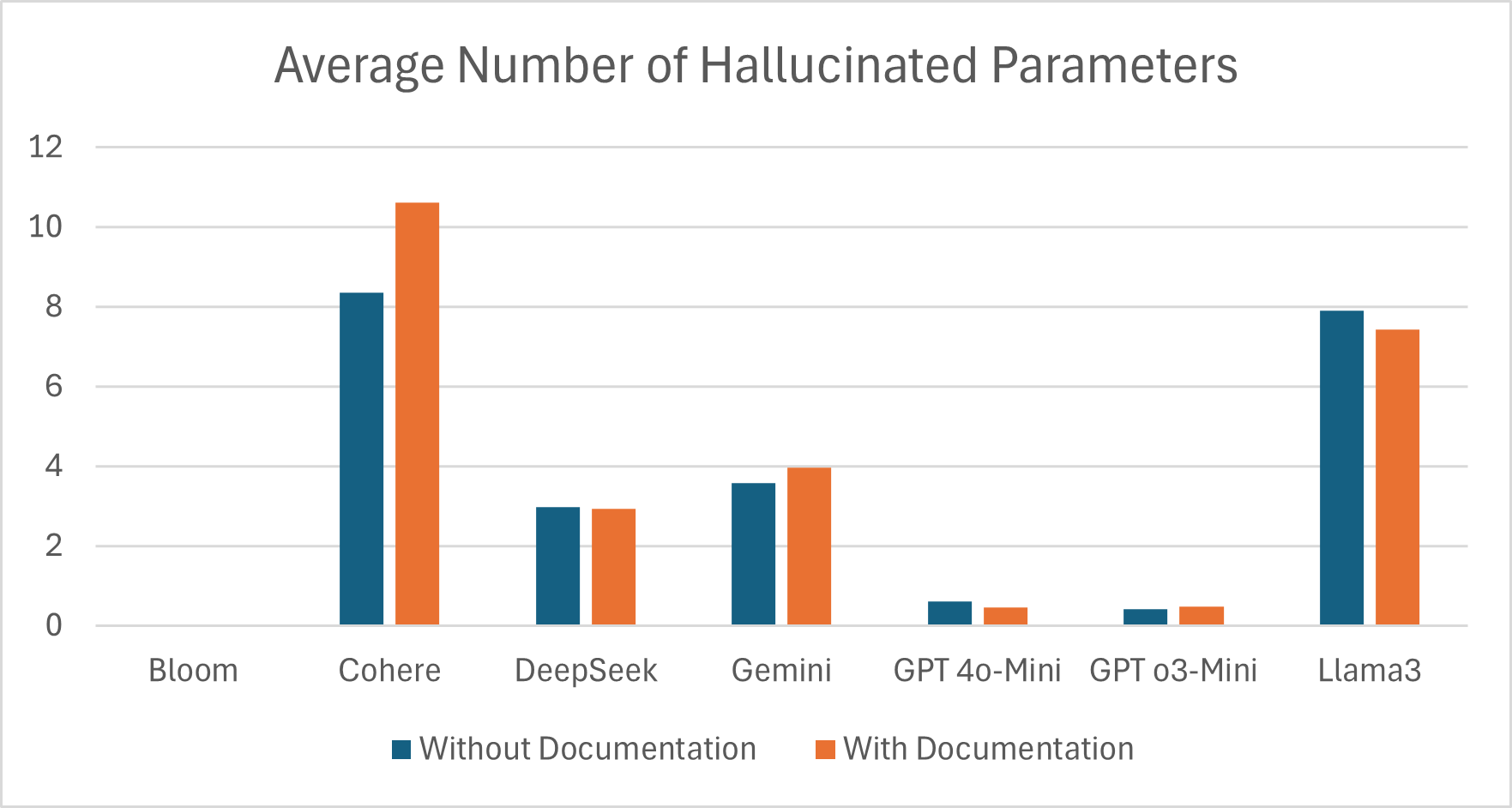}
    \caption{Average Number of Hallucinated Parameters per LLM}
    \label{fig:avg_hparams}
\end{figure}

When asked to generate pwquality.conf files, some LLMs tend to hallucinate parameters that don't exist. Figure \ref{fig:avg_hparams} keeps track of the average number of non-existent parameters an LLM hallucinates per generation. Bloom seems to perform the best, but it's always at 0 due to it not generating anything to begin with. In reality, the best performing are GPT 4o-mini and o3-mini, each with less than one hallucinated parameter on average. Interestingly, providing the documentation helps 4o-mini hallucinate less, but makes o3-mini hallucinate more. Whether the documentation helps or not seems to be LLM dependent, with it slightly reducing hallucination in Deepseek, 4o-mini, and Llama3, while increasing Hallucination in Cohere, Gemini, and o3-mini. Cohere consistently hallucinates the most non-existent parameters, and is also made significantly worse when provided documentation, even more so than the others. 

\subsubsection{Effects of Hallucinations}
While some LLMs hallucinate a lot of parameters, we have noticed that it doesn't affect the generated file when plugged in. If a non-existent parameter like "check\_userpass" is included, the PAM module will simply ignore it and enforce all the valid parameters such as "minlen". This means a file containing all correctly assigned parameters and a few hallucinated ones is functionally the same as one that doesn't contain the hallucinated parameters, hence why we measured consistency both including and excluding hallucinated parameters, and measured accuracy while completely ignoring them.  

\subsection{Outlier Policy Prompts}
Our natural language password prompts contain a mix of enforceable parameters as well as red herrings that cannot be enforced in the context of the pwquality.conf file. Within these prompts, there are two outliers that stick out, listed in Table \ref{tbl:prompts}. The first prompt is the default setting of the pwquality.conf file translated into plain text. This contains no red herrings, and the default parameters don't even have to be changed. To pass this, an LLM could even return a blank file, as a blank file automatically assigns all defaults. In contrast to the first, the second prompt is a large verbose set of instructions mostly comprised of red herrings, with only two policies directly enforceable by the pwquality.conf file, both of which are also set to their default assignments. We have specifically isolated results (prompts not including documentation) from these two extremes to compare the differences in LLM behavior. 

\subsubsection{Consistency}
\begin{figure}[tb]
    \centering
    \includegraphics[width=\linewidth]{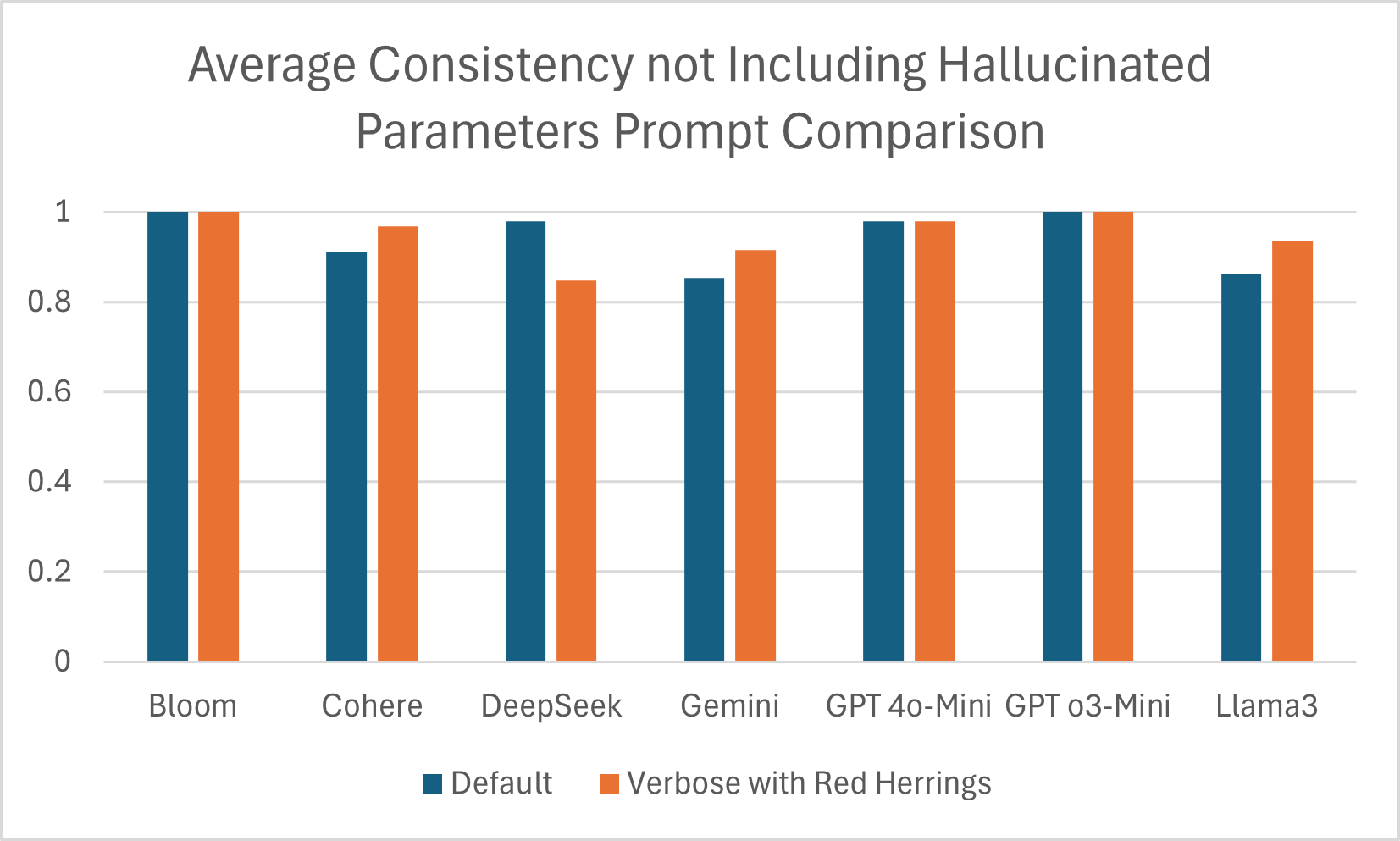}
    \caption{Average Consistency Prompt 4 vs Prompt 5 (no documentation)}
    \label{fig:avg_cons_comp}
\end{figure}
Figure \ref{fig:avg_cons_comp} displays the average consistency for every LLM for the two prompts, only including real parameters and not including documentation. Barring Bloom's blank responses, GPT o3-mini is the only LLM to be 100\% consistent for both prompts. Likely due to the smaller number of assignable parameters, Cohere, Gemini, and Llama3 are more consistent with the red herring prompt, showing that the fluff didn't affect the consistency of generation. The only LLM to see a significant reduction in consistency with the red herring prompt is DeepSeek. Interestingly, DeepSeek's consistency with the default prompt is greater than any of the LLMs which were noticeably more consistent with the red herring prompt. The two OpenAI models do the best on average, and they have a negligible consistency difference between the two prompts. 

\subsubsection{Correctness}

\begin{figure}[tb]
    \centering
    \includegraphics[width=\linewidth]{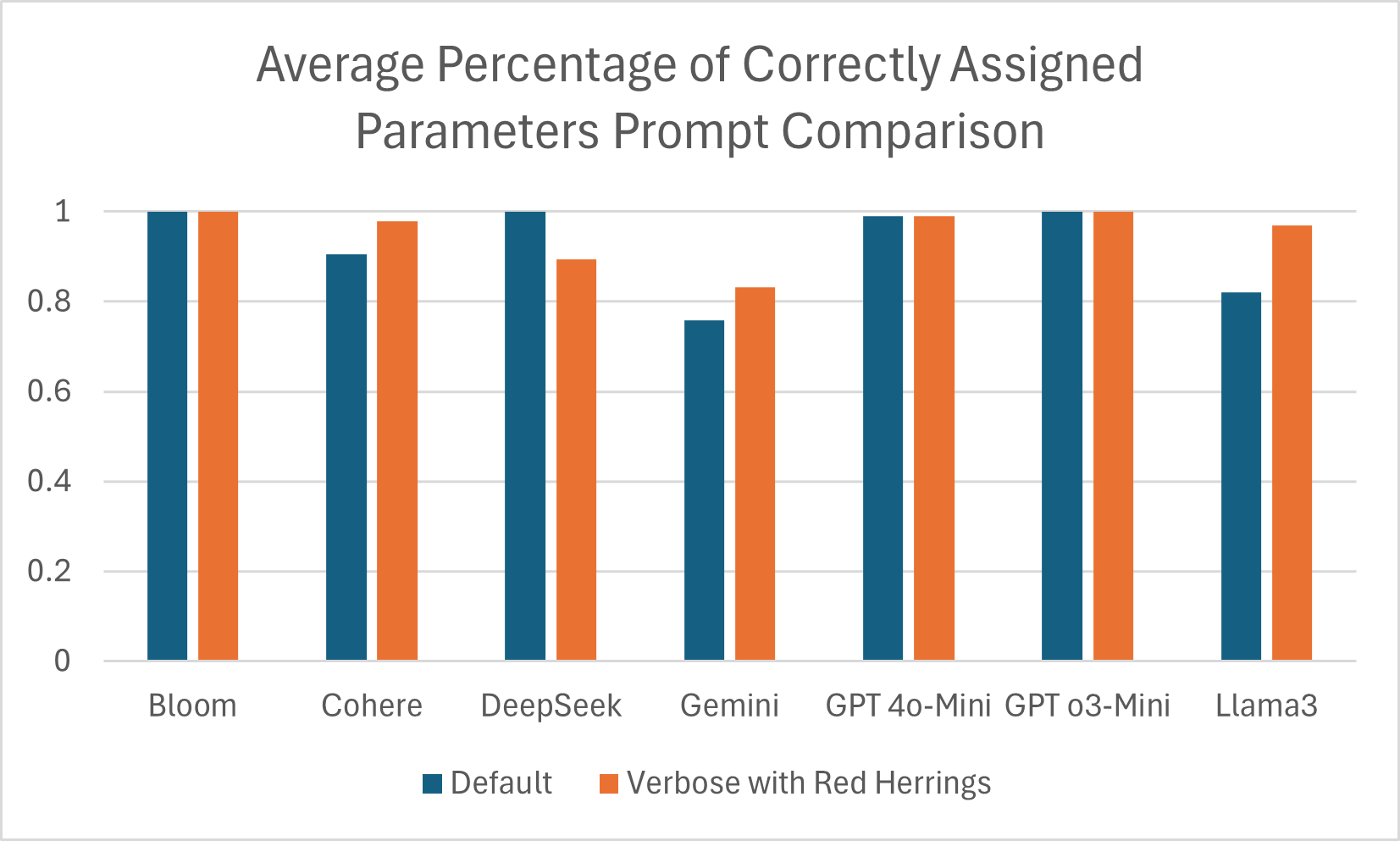}
    \caption{Average Number of Correctly Assigned Parameters per LLM Prompt 4 vs Prompt 5 (no documentation)}
    \label{fig:avg_corr_comp}
\end{figure}

Figure \ref{fig:avg_corr_comp} displays the average correctly assigned parameters for the two prompts, only including real parameters and not including documentation. Bloom again has the full 100\% here since all the parameter assignments here are defaults. Again, the OpenAI models perform the best, with GPT o3-mini doing slightly better than 4o-mini. Correctness displays the exact same trends as consistency in terms of which prompt results in better results. Cohere, Gemini and Llama3 tend to correctly assign parameters with the red herring prompt better than with the default. DeepSeek is the opposite and is considerably more accurate than the three for the default prompt. 

\subsubsection{Hallucinated Parameters}
\begin{figure}[tb]
    \centering
    \includegraphics[width=\linewidth]{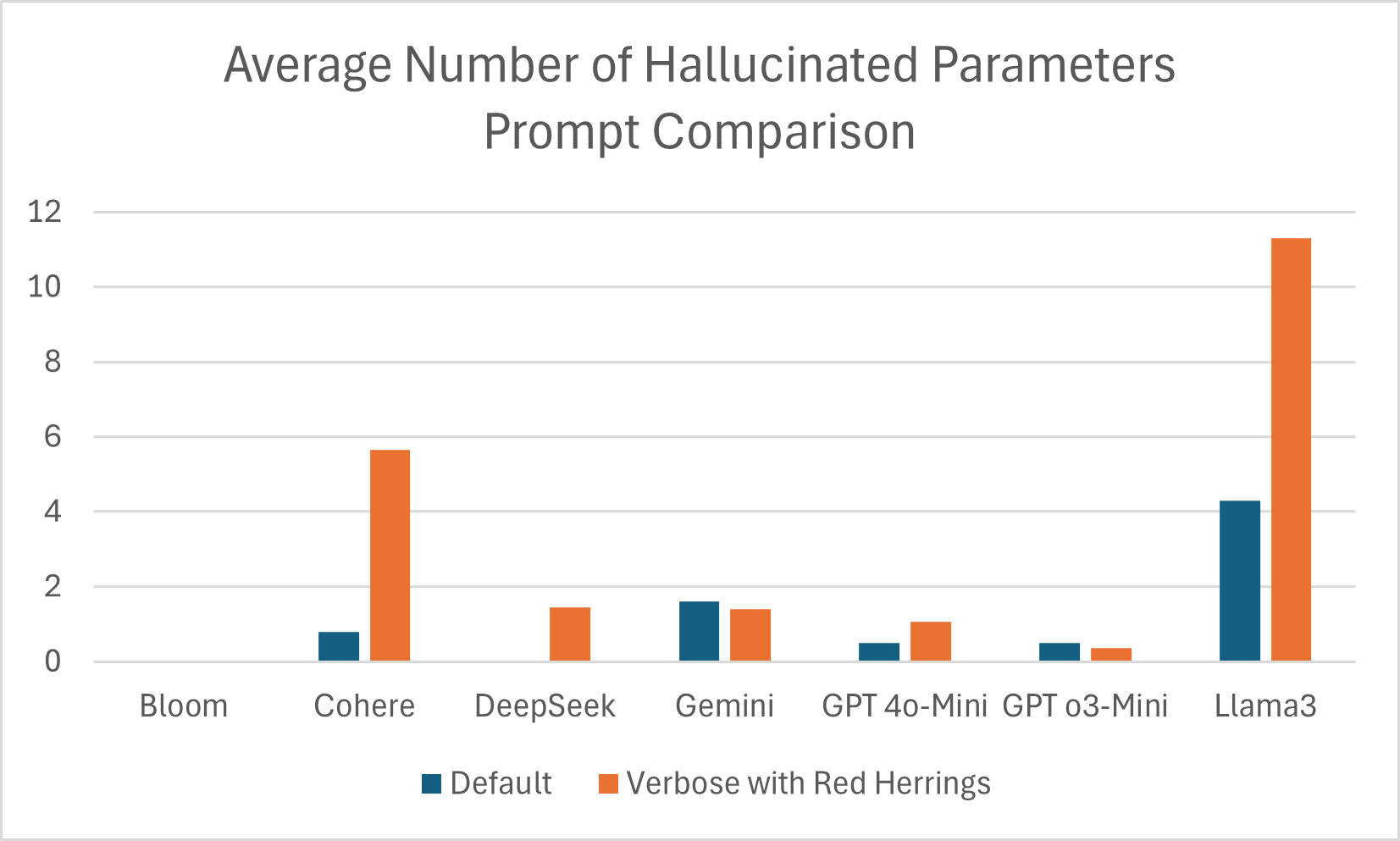}
    \caption{Average Number of Hallucinated Parameters per LLM, Prompt 4 vs Prompt 5 (no documentation)}
    \label{fig:avg_hparams_comp}
\end{figure}
Figure \ref{fig:avg_hparams_comp} displays the average number of hallucinated parameters for the two prompts. As expected, on average, LLMs hallucinate more parameters with the verbose red herring prompt. Llama3 does the worst here, displaying both the largest overall number of hallucinated parameters and the largest difference between the default and the verbose prompt. Cohere is close behind, which is in contrast to the overall average, where Cohere hallucinates the most parameters. Gemini and GPT o3-mini are the outliers, hallucinating slightly more parameters on average with the default prompt. Along with that DeepSeek is the only LLM to never hallucinate a single parameter for one of the prompts, that being the default. This follows its trend of being the outlier that does considerably better with the default prompt. This metric also follows the trend of the OpenAI models doing the best on average. 

\subsection{Limitations and Scope of Automation}
While our work demonstrates that LLMs can be evaluated for configuration generation with quantitative rigor, it is essential to acknowledge the boundary conditions of this automation. The generated configuration file (pwquality.conf) operates at the interface between human policy intent and system-enforced password restrictions. However, the actual enforcement logic resides within the pam\_pwquality.so binary, which is part of the PAM stack and not subject to direct modification through configuration. Moreover, PAM includes an internal failsafe mechanism that overrides insecure or outdated configurations, rendering certain LLM-generated configurations ineffective even if syntactically correct. This highlights a practical constraint: the full behavior of the password policy enforcement depends not only on the generated file, but also on the state and version of the PAM module. Our system does not currently intervene at that layer. Future work may explore automated validation of effective policy behavior by combining config generation with PAM-aware policy simulation or static analysis.

\section{Related Work}\label{sec:related}
Recent advances in large language models (LLMs) have transformed the landscape of natural language processing. Models such as GPT-3.5 and GPT-4 have demonstrated impressive capabilities in generating human-like text, while open-source alternatives like LLaMA and OPT have broadened accessibility and research opportunities~\cite{openai2023gpt35,openai2023gpt4,touvron2023llamaopenefficientfoundation,zhang2022optopenpretrainedtransformer}. These models serve as the backbone for various applications, ranging from everyday language tasks to more specialized domains like cybersecurity.

A critical research focus has been on the consistency of LLM outputs—a key factor in establishing trust and reliability in these systems. For example, Jang and Lukasiewicz analyzed the consistency of ChatGPT responses~\cite{jang2023consistencyanalysischatgpt}, and later work by Jang et al. introduced BECEL and TruthEval, which are benchmarks designed to evaluate different aspects of consistency in LLMs~\cite{jang-etal-2022-becel, khatun2024truthevaldatasetevaluatellm}. Additional studies have expanded on this foundation; Zhu et al.\cite{10651158} and Lee et al.\cite{lee2024evaluatingconsistencyllmevaluators} propose novel frameworks for measuring model consistency, while Ye et al.~\cite{ye2024flask} offer a fine-grained evaluation approach based on alignment skill sets. 

Beyond baseline consistency analyses, recent studies have investigated methods that enhance the reasoning capabilities of LLMs. For instance, techniques like chain-of-thought prompting~\cite{wei2023chainofthoughtpromptingelicitsreasoning} and self-consistency approaches~\cite{wang2023selfconsistency} have demonstrated improvements in multi-step reasoning tasks. Furthermore, advances in prompt engineering, adversarial training, and retrieval-augmented generation (RAG)~\cite{lewis2021retrievalaugmentedgenerationknowledgeintensivenlp} have shown promise in boosting LLM robustness. However, since these techniques are primarily tailored for general-purpose applications, they might not fully capture the specific nuances required for access control systems. Collectively, these approaches not only enhance task accuracy but also provide valuable insights into the variability of model outputs, highlighting the need for a unified framework for consistency evaluation. 

Parallel to these developments, the cybersecurity field is increasingly leveraging LLMs for practical applications in access control and security policy formulation. Secure code copilots and the generation of secure code have emerged as key research areas, addressing challenges in the analysis of code and configurations~\cite{toth2024llms, minna2024analyzing}, secure code generation~\cite{vaidya2023critical,saha2024empowering}, and even security and penetration testing~\cite{song2024poster}. Although research specifically targeting password policy generation is still in its early stages, the integration of LLM capabilities with cybersecurity tasks is already evident. Recent studies have demonstrated that generative AI can significantly bolster cybersecurity operations~\cite{10534674,motlagh2024largelanguagemodelscybersecurity}, while the development of specialized benchmarks and datasets for cybersecurity evaluation~\cite{10679494,yu2025primuspioneeringcollectionopensource} further underscores the potential for automating and enhancing access control systems. 

Our work extends these prior contributions by adapting and enhancing existing consistency evaluation methodologies specifically for access control scenarios. In doing so, we aim to provide a focused analysis of the role that consistency plays in secure, AI-driven access control, and to offer practical insights into mitigating the unique challenges—such as hallucination and adversarial manipulation—that affect policy enforcement in critical cybersecurity environments. 

\section{Conclusions and Future Work}\label{sec:conclusion}

In this study, we presented a formal definition of Soundness tailored to the generation of password policies and introduced a comprehensive framework for evaluating the consistency of LLM outputs. By defining Soundness in terms of functional equivalence, we established a rigorous baseline for LLM-generated pwquality.conf files against manually constructed benchmarks. 

Our evaluation employed two experimental frameworks: one where pre-trained LLMs generated configuration files based solely on natural language prompts, and another that employed the official pwquality.conf documentation. Extensive evaluations on multiple models reveal that, although some LLMs demonstrate high consistency, others exhibit discrepancies and hallucinations that may lead to inconsistent access control enforcement across networks. Our work contributes a crucial step toward understanding and improving the consistency of LLM-generated configurations, offering insights for advancing trustworthy AI in Cybersecurity Applications.


Future research should focus on refining evaluation methodologies, exploring more robust In-Context Learning strategies, and extending the framework to other cybersecurity domains. We intend to do so by exploring other ways of enforcing password policies, such as asking a model to generate files that it deems necessary for the Linux system. We also plan to explore other directions which involve scouring parameters - hallucinated or otherwise - and incorporating them into such files to exploit vulnerabilities in the Linux and other contemporary systems. Traversing such relationships between consistency and hallucinations in detail may help us classify and fine-tune LLMs for different cybersecurity tasks better. 

In addition to refining evaluation metrics, a promising avenue for future work lies in creating a task-specific dataset comprising natural language password policies and their corresponding pwquality.conf implementations. Such a dataset could serve as the basis for fine-tuning LLMs or exploring domain-adaptive techniques such as retrieval-augmented generation (RAG) or instruction tuning. Although this paper focuses on zero-shot prompting to establish baseline performance, our findings suggest that specialized training could substantially improve the accuracy, consistency, and soundness of generated configurations. Beyond pwquality.conf, we envision extending our framework to other configuration files within the Linux authentication system—such as login.defs, system-auth, or even sshd\_config—where LLM-generated misconfigurations may introduce subtle but critical vulnerabilities. This direction holds the potential to transform LLMs into trustworthy assistants for secure, automated system administration.

\bibliographystyle{IEEEtranS}
\bibliography{references}

\begin{thebibliography}{10}
\providecommand{\url}[1]{#1}
\csname url@samestyle\endcsname
\providecommand{\newblock}{\relax}
\providecommand{\bibinfo}[2]{#2}
\providecommand{\BIBentrySTDinterwordspacing}{\spaceskip=0pt\relax}
\providecommand{\BIBentryALTinterwordstretchfactor}{4}
\providecommand{\BIBentryALTinterwordspacing}{\spaceskip=\fontdimen2\font plus
\BIBentryALTinterwordstretchfactor\fontdimen3\font minus \fontdimen4\font\relax}
\providecommand{\BIBforeignlanguage}[2]{{%
\expandafter\ifx\csname l@#1\endcsname\relax
\typeout{** WARNING: IEEEtranS.bst: No hyphenation pattern has been}%
\typeout{** loaded for the language `#1'. Using the pattern for}%
\typeout{** the default language instead.}%
\else
\language=\csname l@#1\endcsname
\fi
#2}}
\providecommand{\BIBdecl}{\relax}
\BIBdecl

\bibitem{cohere2023}
``Cohere api documentation,'' \url{https://cohere.ai/}, 2023.

\bibitem{10534674}
N.~Capodieci, C.~Sanchez-Adames, J.~Harris, and U.~Tatar, ``The impact of generative ai and llms on the cybersecurity profession,'' in \emph{2024 Systems and Information Engineering Design Symposium (SIEDS)}, 2024, pp. 448--453.

\bibitem{cis_password_guide}
\BIBentryALTinterwordspacing
{Center for Internet Security (CIS)}, ``Cis password policy guide: Passphrases, monitoring, and more,'' 2023, accessed: 2025-03-30. [Online]. Available: \url{https://learn.cisecurity.org/cis-password-policy-guide-passphrases-monitoring-and-more}
\BIBentrySTDinterwordspacing

\bibitem{deepseekai2025deepseekv3technicalreport}
\BIBentryALTinterwordspacing
A.~L. et. al, ``Deepseek-v3 technical report,'' 2025. [Online]. Available: \url{https://arxiv.org/abs/2412.19437}
\BIBentrySTDinterwordspacing

\bibitem{jang-etal-2022-becel}
\BIBentryALTinterwordspacing
M.~Jang, D.~S. Kwon, and T.~Lukasiewicz, ``{BECEL}: Benchmark for consistency evaluation of language models,'' in \emph{Proceedings of the 29th International Conference on Computational Linguistics}.\hskip 1em plus 0.5em minus 0.4em\relax Gyeongju, Republic of Korea: International Committee on Computational Linguistics, Oct. 2022, pp. 3680--3696. [Online]. Available: \url{https://aclanthology.org/2022.coling-1.324/}
\BIBentrySTDinterwordspacing

\bibitem{jang2023consistencyanalysischatgpt}
\BIBentryALTinterwordspacing
M.~E. Jang and T.~Lukasiewicz, ``Consistency analysis of chatgpt,'' 2023. [Online]. Available: \url{https://arxiv.org/abs/2303.06273}
\BIBentrySTDinterwordspacing

\bibitem{khatun2024truthevaldatasetevaluatellm}
\BIBentryALTinterwordspacing
A.~Khatun and D.~G. Brown, ``Trutheval: A dataset to evaluate llm truthfulness and reliability,'' 2024. [Online]. Available: \url{https://arxiv.org/abs/2406.01855}
\BIBentrySTDinterwordspacing

\bibitem{lee2024evaluatingconsistencyllmevaluators}
\BIBentryALTinterwordspacing
N.~Lee, J.~Hong, and J.~Thorne, ``Evaluating the consistency of llm evaluators,'' 2024. [Online]. Available: \url{https://arxiv.org/abs/2412.00543}
\BIBentrySTDinterwordspacing

\bibitem{lewis2021retrievalaugmentedgenerationknowledgeintensivenlp}
\BIBentryALTinterwordspacing
P.~Lewis, E.~Perez, A.~Piktus, F.~Petroni, V.~Karpukhin, N.~Goyal, H.~Küttler, M.~Lewis, W.~tau Yih, T.~Rocktäschel, S.~Riedel, and D.~Kiela, ``Retrieval-augmented generation for knowledge-intensive nlp tasks,'' 2021. [Online]. Available: \url{https://arxiv.org/abs/2005.11401}
\BIBentrySTDinterwordspacing

\bibitem{arch_pwquality}
\BIBentryALTinterwordspacing
A.~Linux, \emph{pwquality.conf(5) — Arch Manual Pages}, n.d., accessed: 2025-03-28. [Online]. Available: \url{https://man.archlinux.org/man/pwquality.conf.5.en}
\BIBentrySTDinterwordspacing

\bibitem{linux_password_defaults}
{Linux Manual}, ``Default password policy in linux systems (e.g., pam, /etc/login.defs),'' 2024, based on standard Linux configurations; no official central documentation.

\bibitem{love2005selinux}
F.~M. Love, K.~M. Anderson, and D.~Smalley, \emph{SELinux by Example: Using Security Enhanced Linux}.\hskip 1em plus 0.5em minus 0.4em\relax Upper Saddle River, NJ, USA: Prentice Hall PTR, 2006.

\bibitem{lumificyber_passwords}
\BIBentryALTinterwordspacing
{Lumi Cyber}, ``Successful password policies for organizations,'' 2023, accessed: 2025-03-30. [Online]. Available: \url{https://www.lumificyber.com/blog/successful-password-policies-for-organizations/}
\BIBentrySTDinterwordspacing

\bibitem{mcdonald2024reducing}
D.~McDonald, R.~Papadopoulos, and L.~Benningfield, ``Reducing llm hallucination using knowledge distillation: A case study with mistral large and mmlu benchmark,'' \emph{Authorea Preprints}, 2024.

\bibitem{minna2024analyzing}
F.~Minna, F.~Massacci, and K.~Tuma, ``Analyzing and mitigating (with llms) the security misconfigurations of helm charts from artifact hub,'' \emph{arXiv preprint arXiv:2403.09537}, 2024.

\bibitem{motlagh2024largelanguagemodelscybersecurity}
\BIBentryALTinterwordspacing
F.~N. Motlagh, M.~Hajizadeh, M.~Majd, P.~Najafi, F.~Cheng, and C.~Meinel, ``Large language models in cybersecurity: State-of-the-art,'' 2024. [Online]. Available: \url{https://arxiv.org/abs/2402.00891}
\BIBentrySTDinterwordspacing

\bibitem{nist_80063b}
\BIBentryALTinterwordspacing
{NIST}, ``Digital identity guidelines: Authentication and lifecycle management (sp 800-63b),'' 2020, accessed: 2025-03-30. [Online]. Available: \url{https://nvlpubs.nist.gov/nistpubs/SpecialPublications/NIST.SP.800-63b.pdf}
\BIBentrySTDinterwordspacing

\bibitem{openai2022chatgpt}
OpenAI, ``Chatgpt: Optimizing language models for dialogue,'' \url{https://openai.com/blog/chatgpt}, 2022, accessed: 2025-03-30.

\bibitem{openai2024gpt4omini}
\BIBentryALTinterwordspacing
{OpenAI}, ``{GPT-4o Mini: Advancing Cost-Efficient Intelligence},'' Jul. 2024. [Online]. Available: \url{https://openai.com/index/gpt-4o-mini-advancing-cost-efficient-intelligence/}
\BIBentrySTDinterwordspacing

\bibitem{openai_o3mini}
\BIBentryALTinterwordspacing
OpenAI, ``o3-mini in chatgpt - faq,'' 2025, accessed March 31, 2025. [Online]. Available: \url{https://help.openai.com/en/articles/10491870-o3-mini-in-chatgpt-faq}
\BIBentrySTDinterwordspacing

\bibitem{openai2023gpt35}
\BIBentryALTinterwordspacing
{OpenAI Team}, ``{GPT-3.5: Generative Pre-trained Transformer},'' OpenAI API, 2023, accessed: Aug. 4, 2024. [Online]. Available: \url{https://platform.openai.com/docs/models/gpt-3-5}
\BIBentrySTDinterwordspacing

\bibitem{patwardhan2024automated}
A.~Patwardhan, V.~Vaidya, and A.~Kundu, ``Automated consistency analysis of llms,'' in \emph{2024 IEEE 6th International Conference on Trust, Privacy and Security in Intelligent Systems, and Applications (TPS-ISA)}.\hskip 1em plus 0.5em minus 0.4em\relax IEEE, 2024, pp. 118--127.

\bibitem{purplesec_passwords}
\BIBentryALTinterwordspacing
{Purplesec}, ``Cyber security policy templates: Password security,'' 2020, accessed: 2025-03-30. [Online]. Available: \url{https://purplesec.us/resources/cyber-security-policy-templates/password-security/}
\BIBentrySTDinterwordspacing

\bibitem{saha2024empowering}
D.~Saha, K.~Yahyaei, S.~K. Saha, M.~Tehranipoor, and F.~Farahmandi, ``Empowering hardware security with llm: The development of a vulnerable hardware database,'' in \emph{2024 IEEE International Symposium on Hardware Oriented Security and Trust (HOST)}.\hskip 1em plus 0.5em minus 0.4em\relax IEEE, 2024, pp. 233--243.

\bibitem{securden_passwords}
\BIBentryALTinterwordspacing
{Securden}, ``Top 10 password policy best practices,'' 2023, accessed: 2025-03-30. [Online]. Available: \url{https://www.securden.com/blog/top-10-password-policies.html}
\BIBentrySTDinterwordspacing

\bibitem{song2024poster}
S.~Song, A.~Kundu, and B.~Tak, ``Poster: Seccomp profiling with dynamic analysis via chatgpt-assisted test code generation,'' in \emph{Proceedings of the 19th ACM Asia Conference on Computer and Communications Security}, 2024, pp. 1928--1930.

\bibitem{geminiteam2024geminifamilyhighlycapable}
\BIBentryALTinterwordspacing
G.~Team and R.~Anil~et. al., ``Gemini: A family of highly capable multimodal models,'' 2024. [Online]. Available: \url{https://arxiv.org/abs/2312.11805}
\BIBentrySTDinterwordspacing

\bibitem{openai2023gpt4}
O.~Team, ``Gpt-4: Generative pre-trained transformer,'' \texttt{OpenAI API}, 2023, \url{https://platform.openai.com/docs/models/gpt-4}.

\bibitem{10679494}
N.~Tihanyi, M.~A. Ferrag, R.~Jain, T.~Bisztray, and M.~Debbah, ``Cybermetric: A benchmark dataset based on retrieval-augmented generation for evaluating llms in cybersecurity knowledge,'' in \emph{2024 IEEE International Conference on Cyber Security and Resilience (CSR)}, 2024, pp. 296--302.

\bibitem{toth2024llms}
R.~T{\'o}th, T.~Bisztray, and L.~Erdodi, ``Llms in web-development: Evaluating llm-generated php code unveiling vulnerabilities and limitations,'' \emph{arXiv preprint arXiv:2404.14459}, 2024.

\bibitem{touvron2023llamaopenefficientfoundation}
\BIBentryALTinterwordspacing
H.~Touvron, T.~Lavril, G.~Izacard, X.~Martinet, M.-A. Lachaux, T.~Lacroix, B.~Rozière, N.~Goyal, E.~Hambro, F.~Azhar, A.~Rodriguez, A.~Joulin, E.~Grave, and G.~Lample, ``Llama: Open and efficient foundation language models,'' 2023. [Online]. Available: \url{https://arxiv.org/abs/2302.13971}
\BIBentrySTDinterwordspacing

\bibitem{vaidya2023critical}
J.~Vaidya and H.~Asif, ``A critical look at ai-generate software: Coding with the new ai tools is both irresistible and dangerous,'' \emph{Ieee Spectrum}, vol.~60, no.~7, pp. 34--39, 2023.

\bibitem{wang2023selfconsistency}
\BIBentryALTinterwordspacing
X.~Wang, J.~Wei, D.~Schuurmans, Q.~V. Le, E.~H. Chi, S.~Narang, A.~Chowdhery, and D.~Zhou, ``Self-consistency improves chain of thought reasoning in language models,'' in \emph{The Eleventh International Conference on Learning Representations}, 2023. [Online]. Available: \url{https://openreview.net/forum?id=1PL1NIMMrw}
\BIBentrySTDinterwordspacing

\bibitem{wei2023chainofthoughtpromptingelicitsreasoning}
\BIBentryALTinterwordspacing
J.~Wei, X.~Wang, D.~Schuurmans, M.~Bosma, B.~Ichter, F.~Xia, E.~Chi, Q.~Le, and D.~Zhou, ``Chain-of-thought prompting elicits reasoning in large language models,'' 2023. [Online]. Available: \url{https://arxiv.org/abs/2201.11903}
\BIBentrySTDinterwordspacing

\bibitem{workshop2023bloom176bparameteropenaccessmultilingual}
\BIBentryALTinterwordspacing
B.~Workshop, :, and T.~L.~S. et. al., ``Bloom: A 176b-parameter open-access multilingual language model,'' 2023. [Online]. Available: \url{https://arxiv.org/abs/2211.05100}
\BIBentrySTDinterwordspacing

\bibitem{ye2024flask}
\BIBentryALTinterwordspacing
S.~Ye, D.~Kim, S.~Kim, H.~Hwang, S.~Kim, Y.~Jo, J.~Thorne, J.~Kim, and M.~Seo, ``{FLASK}: Fine-grained language model evaluation based on alignment skill sets,'' in \emph{The Twelfth International Conference on Learning Representations}, 2024. [Online]. Available: \url{https://openreview.net/forum?id=CYmF38ysDa}
\BIBentrySTDinterwordspacing

\bibitem{yu2025primuspioneeringcollectionopensource}
\BIBentryALTinterwordspacing
Y.-C. Yu, T.-H. Chiang, C.-W. Tsai, C.-M. Huang, and W.-K. Tsao, ``Primus: A pioneering collection of open-source datasets for cybersecurity llm training,'' 2025. [Online]. Available: \url{https://arxiv.org/abs/2502.11191}
\BIBentrySTDinterwordspacing

\bibitem{zhang2022optopenpretrainedtransformer}
\BIBentryALTinterwordspacing
S.~Zhang, S.~Roller, N.~Goyal, M.~Artetxe, M.~Chen, S.~Chen, C.~Dewan, M.~Diab, X.~Li, X.~V. Lin, T.~Mihaylov, M.~Ott, S.~Shleifer, K.~Shuster, D.~Simig, P.~S. Koura, A.~Sridhar, T.~Wang, and L.~Zettlemoyer, ``Opt: Open pre-trained transformer language models,'' 2022. [Online]. Available: \url{https://arxiv.org/abs/2205.01068}
\BIBentrySTDinterwordspacing

\bibitem{zhu2021apparmor}
H.~Zhu and C.~Gehrmann, ``Apparmor profile generator as a cloud service,'' in \emph{Proceedings of the 11th International Conference on Cloud Computing and Services Science}.\hskip 1em plus 0.5em minus 0.4em\relax SciTePress, 2021, pp. 45--55.

\bibitem{10651158}
Q.~Zhu, D.~Lyu, X.~Fan, X.~Wang, Q.~Tu, Y.~Zhan, and H.~Chen, ``Multi-model consistency for llms’ evaluation,'' in \emph{2024 International Joint Conference on Neural Networks (IJCNN)}, 2024, pp. 1--8.

\end{thebibliography}
\vspace{12pt}

\clearpage
\onecolumn
\appendix
\begin{table}[h]
\caption{Password Policy Requirements}
\label{tbl:prompts}
\vspace{0.5em}
\renewcommand{\arraystretch}{1.4}
\begin{tabular}{|p{3cm}|p{14.5cm}|}
\hline
\textbf{Policy} & \textbf{Requirements} \\
P1 & There must be at least 1 character in the new password that isn't in the old password. \newline
    The password must be at least 8 characters long. \newline
    Cannot contain words in the cracklib directory. \newline
    Cannot contain the username in any form. \newline
    Must be enforced by the PAM module. \newline
    Allows the user 3 retries before returning an error. \\
\hline
P2 & A Memorized Secret authenticator — commonly referred to as a password or, if numeric, a PIN — is a secret value intended to be chosen and memorized by the user. Memorized secrets need to be of sufficient complexity and secrecy that it would be impractical for an attacker to guess or otherwise discover the correct secret value. A memorized secret is something you know. \newline
Memorized secrets SHALL be at least 8 characters in length if chosen by the subscriber. Memorized secrets chosen randomly by the CSP or verifier SHALL be at least 6 characters in length and MAY be entirely numeric. If the CSP or verifier disallows a chosen memorized secret based on its appearance on a blacklist of compromised values, the subscriber SHALL be required to choose a different memorized secret. No other complexity requirements for memorized secrets SHOULD be imposed. \newline
Verifiers SHALL require subscriber-chosen memorized secrets to be at least 8 characters in length. Verifiers SHOULD permit subscriber-chosen memorized secrets at least 64 characters in length. All printing ASCII characters, as well as the space character, SHOULD be acceptable in memorized secrets. Unicode characters SHOULD be accepted as well. \newline
To account for mistyping, verifiers MAY replace multiple consecutive space characters with a single space prior to verification, provided the result is still 8+ characters. Truncation SHALL NOT be performed. Unicode secrets SHOULD use the NFKC or NFKD normalization process before hashing. \newline
Verifiers SHALL NOT permit storage of password “hints” accessible to unauthenticated users, nor prompt for specific types of personal information (e.g., "first pet"). \newline
When processing password changes, secrets SHALL be compared to a list of known weak/compromised values. These include breached passwords, dictionary words, sequential patterns (e.g., "aaaaaa", "123abc"), and context-specific tokens like the username or service name. If found on the list, the user SHALL be prompted to choose a new secret. \newline
Verifiers SHOULD offer feedback tools like a password-strength meter. \newline
Rate limiting SHALL be applied to failed login attempts. Arbitrary password expiration SHOULD NOT be required unless there is evidence of compromise. \\
\hline
\end{tabular}
\end{table}

\end{document}